\documentclass[sigconf,edbt]{acmart-edbt2024}

\def\BibTeX{{\rm B\kern-.05em{\sc i\kern-.025em b}\kern-.08em
    T\kern-.1667em\lower.7ex\hbox{E}\kern-.125emX}}

\usepackage{booktabs} %
\usepackage{enumitem}
\usepackage[utf8]{inputenc}
\usepackage[listings,skins,breakable]{tcolorbox}
\usepackage[T1]{fontenc}
\usepackage{microtype}
\usepackage[a-2b]{pdfx}
\usepackage{graphicx}
\usepackage{balance} 

\usepackage{subfigure}
\usepackage{multirow}
\usepackage{algorithm}
\usepackage{algorithmic}
\usepackage{color}
\usepackage{listings}
\usepackage{amsmath}
\usepackage{mathtools}
\usepackage{float}

\usepackage{tabularx}
\usepackage{amsmath}
\usepackage{subscript}
\usepackage[skip=0pt]{caption}
\usepackage[labelfont=bf]{caption}

\newcommand{\eat}[1]{}

\setcopyright{rightsretained}

\acmDOI{}

\acmISBN{978-3-89318-095-0}

\acmConference[EDBT 2024]{27th International Conference on Extending Database Technology (EDBT)}{25th March-28th March, 2024}{Paestum, Italy}
\acmYear{2024}

\begin{document}
\title{Serving Deep Learning Models from Relational Databases}

\author{Lixi Zhou$^a$, Qi Lin$^a$, Kanchan Chowdhury$^a$, Saif Masood$^a$, \\ Alexandre Eichenberger$^{c*}$, Hong Min$^{c*}$, Alexander Sim$^{e*}$,  Jie Wang$^{b*}$, \\ Yida Wang$^{b*}$, Kesheng Wu$^{e*}$, Binhang Yuan$^{d*}$,  Jia Zou$^a$ \thanks{* These authors are ordered alphabetically; Jia Zou (jia.zou@asu.edu) is the corresponding author.}}

\affiliation{
  \institution{$^a$ Arizona State University, $^b$ Amazon, $^c$ IBM T. J. Watson Research Center, \\$^d$ Hong Kong University of Science and Technology, $^e$ Lawrence Berkeley National Lab}
  \country{}
  }

\renewcommand{\shortauthors}{zhou et al.}

\begin{abstract}
Serving deep learning (DL) models on relational data has become a critical requirement across diverse commercial and scientific domains, sparking growing interest recently. In this visionary paper, we embark on a comprehensive exploration of representative architectures to address the requirement. We highlight three pivotal paradigms: The state-of-the-art \textit{DL-centric} architecture offloads DL computations to dedicated DL frameworks. The potential \textit{UDF-centric} architecture encapsulates one or more tensor computations into User Defined Functions (UDFs) within the relational database management system (RDBMS). 
The potential \textit{relation-centric} architecture aims to represent a large-scale tensor computation through relational operators. While each of these architectures demonstrates promise in specific use scenarios, we identify urgent requirements for seamless integration of these architectures and the middle ground in-between these architectures. We delve into the gaps that impede the integration and explore innovative strategies to close them. We present a pathway to establish a novel RDBMS for enabling a broad class of data-intensive DL inference applications.
\end{abstract}

\maketitle

\section{Introduction}
\label{sec:intro}

Recently, key applications emerged from the desire to nest SQL queries with deep learning inferences, including but not limited to the following categories:

\noindent
$\bullet$ \textbf{Commercial applications} such as credit card fraud detection, personalized recommendation, customer service chatbots, and anti-money laundering (AML). These use cases increasingly rely on deep learning~\cite{fraud-ibm, cheng2020spatio, xiang2023semi, naumov2019deep, gao2018neural, breslow2017new, creditcardAI}, and are latency-critical. In these cases, transaction data, order data, and customer profiles are usually managed by a relational database management system (RDBMS), and subject to SQL queries for updates and analytics. 

\noindent
$\bullet$ \textbf{Scientific applications} also benefit from the intriguing capabilities of integrating deep learning and RDBMS for efficient surrogate models and self-driving lab experiments\cite{ratner2019office}. For example, real-time control coupled with material modeling requires fast access to pre-computed results to trigger model inferences.~\cite{ratner2019office, macleod2020self}.

\noindent
$\bullet$ \textbf{AI-enhanced autonomous {RDBMS}} %
 increasingly uses Deep Learning (DL)~\cite{yuan2020automatic, kim2022learned, kipf2018learned, zou2021lachesis, zhou2023deepmapping}. In order to minimize the runtime overheads incurred by AI techniques, it is critical to reduce the latency of DL inferences for these techniques.

The DL models in these workloads range from simple models such as feed-forward neural networks (FFNN) and convolutional neural networks (CNN) to complex models such as Transformer~\cite{huang2010hibench} and recommendation models~\cite{naumov2019deep}, as well as the foundation models~\cite{bommasani2021opportunities}, including large language models (LLMs)~\cite{peng2023instruction}. 

\begin{figure*}[h]
\vspace{-5pt}
\centering{%
   \includegraphics[width=7in]{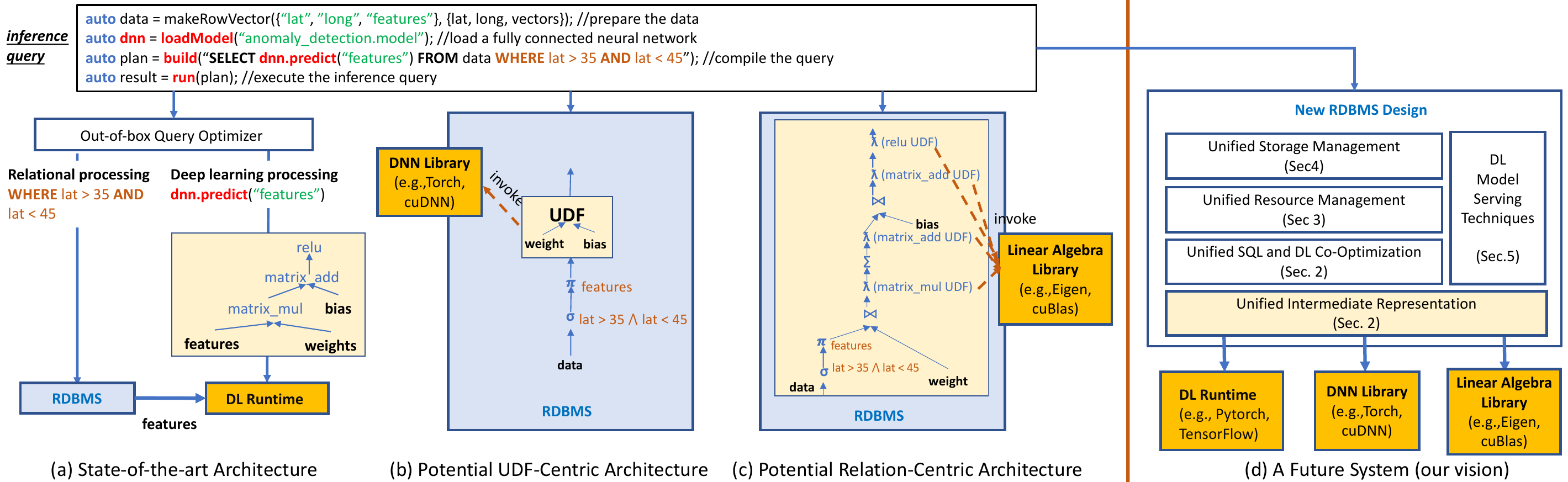}  
}
\vspace{-10pt}
\caption{\small\label{fig:representation-overview} Overview of the three pivotal architectures and our vision  }
\end{figure*}

The state-of-the-art architecture for supporting inference queries, termed \textbf{the DL-centric architecture},  offloads the inference computations to decoupled DL runtimes, as illustrated in Fig.~\ref{fig:representation-overview}a. For example, Amazon Redshift offloads the inference to SageMaker, Microsoft Raven~\cite{karanasos2019extending, park2022end} offloads the inference from Microsoft SQLServer to the ONNX runtime~\cite{onnx}, Google BigQuery offloads the inference to TensorFlow~\cite{tensorflow}, PostgresML~\cite{postgresml} offloads the inference of language models to HuggingFace~\cite{huggingface}. In addition, the emerging EvaDB~\cite{evadb} also runs the inference computations nested in SQL queries in decoupled ML systems such as Hugging Face, OpenAI, YOLO, Stable Diffusion, etc. %

In addition, there are two potential architectures as alternatives, which have not gained enough attention yet.

 \noindent
$\bullet$ The potential \textbf{UDF-centric architecture} encapsulates inference logic as UDFs within RDBMS, as illustrated in Fig.~\ref{fig:representation-overview}b.  However, existing systems such as VerticaML~\cite{fard2020vertica} and PostgresML~\cite{postgresml} mostly use UDF-encapsulation for traditional ML such as XGBoost and logistic regression. There are few existing implementations of UDF-centric systems for DL, although DL offers diverse libraries on CPU/GPU~\cite{cudnn, zdnn, Eigen} for implementing the UDFs. %

 \noindent
$\bullet$ The potential \textbf{relation-centric architecture}, as illustrated in Fig.~\ref{fig:representation-overview}c, represents a model parameter tensor as a relation, i.e., a collection of tensor blocks, and breaks down a tensor operator into multiple relational operators that nest with fine-grained UDFs. This approach has been advocated by SystemsML~\cite{boehm2016systemml} and Tensor Relational Algebra ~\cite{DBLP:journals/pvldb/YuanJZTBJ21, luo2018scalable, jankov12declarative} for scaling to massive tensor operations. However, this architecture hasn't been widely adopted for serving state-of-the-art DL models.

\vspace{3pt}
\noindent
\textbf{Why do we need a new architecture? } In this vision paper, we argue for a new architecture because of the significant shortcomings that we observed in all existing architectures:

\noindent
{$\bullet$  The DL-centric architecture leads to significant cross-system overheads since features that were prepared by the data processing system need to be transferred to separate ML systems for inferences, which renders them inappropriate for latency-critical applications. In addition, the decoupled architecture prevents the inference computations and the relational query processing from being fully co-optimized. Moreover, handling large tensors often incurs out-of-memory (OOM) errors in resource-constraint environments. Finally, managing multiple systems incurs high operational costs.}

\noindent
{$\bullet$ The UDF-centric architecture highly relies on the efficiency of the underlying libraries. It lacks the flexibility in optimizing and scheduling the operations within the UDF in a fine-grained style. As a result, it shares many problems with the DL-centric architecture, such as the difficulty in co-optimizing the relational processing and UDF processing and in handling large-scale tensors. %

\noindent
{$\bullet$ The relation-centric architecture facilitates co-optimization of inference computations and relational processing by converting the former to the latter. In addition, it views a tensor as a collection of tensor blocks, which can resolve the OOM errors in resource-constrained CPU/GPU environments leveraging RDBMS's disk spilling and caching capabilities. However, not all DL operations can benefit from the conversion into relational processing. It significantly increases latency by converting small model inference computations that fit into the CPU/GPU caches from UDF processing into relational processing. }

{On one hand, each existing architecture has significant limitations to meet the latency and resource requirements of various applications. On the other hand, it is challenging for users to identify the optimal architectural design and it is also not ideal to develop a domain-specific solution for each type of application, which requires significant systems expertise and expensive development costs. Therefore, there exists an urgent requirement for a new architecture with a user-friendly interface and an efficient and intelligent kernel that is adaptive to a diverse array of applications.}

\eat{

}

\vspace{3pt}
\noindent
{\textbf{Our Vision:} 
In this work, we advocate for a novel 
RDBMS %
that seamlessly integrates the DL-centric, UDF-centric, and relation-centric architectures as well as various middle grounds in between these architectures. The envisioned system consists of the following novel components, as illustrated in Fig.~\ref{fig:representation-overview}:
}

\noindent
{$\bullet$ A novel unified intermediate representation (IR) to enable a novel query optimizer that not only dynamically selects one of the DL-centric, UDF-centric, and relation-centric representations for each operator, but also co-optimizes the SQL processing and the model inferences by designing novel transformation rules (Sec.~\ref{sec:cross-optimization}). }

\noindent
{$\bullet$  A novel unified resource management framework for tuning the threading and memory allocation for DB operations, DL runtime operations, DNN library operations, and linear algebra operations, and dispatching these operations to devices (Sec.~\ref{sec:resources}).}

\noindent
{$\bullet$ A novel storage co-optimization framework for tensor blocks and relational data that facilitates new techniques, such as the novel accuracy-aware data/model deduplication and the novel physics database design that considers data/model co-partitioning (Sec.~\ref{sec:storage}).
}

\noindent
{$\bullet$ Model serving techniques newly adapted for RDBMS, such as using RDBMS indexing to cache inference results in an application-aware style (Sec.~\ref{sec:serving}). }

\noindent
 \textbf{Expected Benefits.} The proposed novel RDBMS %
 design has numerous advantages over the state-of-the-art architecture and other potential architectures. First, it enhances the productivity of developing applications that desire DB functionalities such as SQL, transaction management, and access control and shortens the time to market. Second, it delivers superior performance for a broad class of inference applications at different scales through unified optimization for computation, resource management, and data storage. Third, it effectively avoids the cross-system overheads for applications bottlenecked by data transfer.  We discussed more benefits such as extensions to training in Sec.~\ref{sec:discussion}. We further validates some of the ideas in Sec.~\ref{sec:evaluation}. %

\vspace{-10pt}
\section{Query Compilation and Optimization} 
\label{sec:cross-optimization}

\noindent
\textbf{Challenge 1.}  How to design a query compilation framework to dynamically generate the DL-centric, UDF-centric, and relation-centric representations for each operator and facilitate co-optimization of relational processing and DL inference computations?

\noindent
\textbf{(1) Unified intermediate representations(IR).}
We consider the following options for designing a unified IR to address the challenge.

\noindent
$\bullet$ Multi-level IR.  
Multi-Level Intermediate Representation (MLIR)~\cite{mlir, lattner2021mlir} is developed by Google and LLVM for integrating multiple levels of IR dialects for machine learning processing. 
Recently, LingoDB~\cite{jungmair2022designing} and Daphne IR~\cite{damme2022daphne} provide an MLIR dialect for relational processing, which can serve as a nice foundation for integrating with existing dialects for AI/ML computations developed by the MLIR community to facilitate cross-optimizations within the MLIR framework. However, a challenging problem is how to co-optimize the DB dialect and ML dialects, and how to glue MLIR and RDBMS runtime for parallel query execution, caching, and indexing~\cite{neumann2011efficiently, neumann2021evolution}. These are not discussed in existing works.

\noindent
\textbf{$\bullet$ A Novel Adaptive In-database IR}. An alternative is to extend existing relational algebra IR to represent linear algebra computations as UDFs or relational operators adaptively. This approach may seamlessly integrate with the RDBMS runtime. For example, we consider an IR, where each node is a traditional relational operator or a model UDF operator. The latter corresponds to a DL model inference computation. Such computation can be lowered to a graph IR, where each node represents a linear algebra operator such as matrix multiplication, matrix addition, relu, softmax, conv2d, etc. 

Both options can support flexible IR transformations required by the envisioned system. First, each linear algebra operator can be flexibly converted into relational operators when a tensor exceeds available memory. For example, a large-scale matrix multiplication can be replaced by a join followed by an aggregation as illustrated in Fig.~\ref{fig:representation-overview}c~\cite{DBLP:journals/pvldb/YuanJZTBJ21}. Second, one or more operators can be offloaded to external DL runtimes, e.g., for pipelining across multiple GPU devices. Third, multiple simple linear algebra operators can be merged into one coarse-grained UDF by invoking libraries, such as CuDNN~\cite{cudnn}, and IBM zDNN~\cite{zdnn}, or using code generation~\cite{chen2018tvm, ragan2013halide, vasilache2018tensor}, to reduce data transfer latency. 

\noindent
\textbf{(2) Co-Optimization.}
A unified IR will open opportunities for co-optimizing SQL processing and DL models, such as novel query transformation rules. For example, consider a pipeline that first joins two separate wide feature datasets $D_1$ and $D_2$ to obtain the entire features and run a FFNN model on top of the features. If the first layer of the model is an embedding layer or a fully connected layer that will reduce the feature dimensions significantly, we may decompose the corresponding weight matrix $W$ into two submatrices $W_1$ and $W_2$, each of which only processes the features belonging to $D_1$ or $D_2$ respectively, so that $W \times (D_1 \bowtie D_2) = (W_1 \times D_1) \bowtie (W_2\times D_2)$. Then, a novel query transformation rule is to push down the submatrix multiplication $W_1 \times D_1$ and $W_2\times D_2$ before we perform the join. It will reduce the intermediate data size if the number of features in the outputs from $W_1\times D_1$ and $W_2 \times D_2$ are significantly smaller than the total number of features in $D_1$ and $D_2$.

Considering the tremendous flexibility, optimizing for complicated inference queries could be expensive. In addition to employing standard techniques such as learning-based optimization~\cite{kim2022learned},  Bayesian optimization~\cite{frazier2018tutorial}, and query graph partitioning~\cite{jankov12declarative}, a promising technique is ahead-of-time (AoT) compilation~\cite{zhang2021adaptive, lee2018pretzel, zheng2022dietcode}. For example, when loading a model into RDBMS, the system will generate multiple execution plans at compilation time and select the best plan at runtime. %

\vspace{5pt}
\section{Unified Resource Management}\label{sec:resources}

\noindent
\textbf{Challenge 2.} How to unify the resource management for RDBMS, DL runtimes, and DL libraries invoked from UDFs?

\noindent
\textbf{(1) Hyper-Parameter Tuning.}
Tuning parameters is important in coordinating resources between the {RDBMS}
 runtime and external DL runtimes colocating on the same machine. For example, 
while configuring the RDBMS buffer pool sizes, we shall also consider the memory requirements of the DL runtimes. Similarly, in the UDF-centric approach, the UDF, e.g., invoking OpenBLAS libraries, may rely on OpenMP for parallelism. We must carefully configure the number of threads for the SQL query processing and OpenMP. Otherwise, significant context switch overheads may occur.  For example, multiple RDBMS
 threads execute the same pipeline stage that contains linear algebra operators in data-parallel style. However, each linear algebra operator may run with a different number of OpenMP threads.

Existing RDBMS
 and ML hyper-parameter tuning works~\cite{butrovich2022tastes, ma2018query, zhou2020database, zhang2019end, li2019qtune, li2017hyperband, snoek2012practical} 
did not consider the heterogeneous nature of these threading configurations of UDF-based linear algebra operators. The heterogeneity in the tasks executed by the RDBMS threads and ML threads made the formulation of the hyper-parameter co-optimization problem challenging. In addition, each ML thread may have a different lifetime, which further complicates the problem. In addition, hyper-parameter search usually requires either significant search latency at the online stage or significant offline overheads for training surrogate models, reinforcement learning, etc. 
Motivated by these challenges, it is promising to explore novel data-efficient hyper-parameter learning techniques, such as meta-learning, zero-shot, and few-shot learning with generative models~\cite{nalisnick2018deep}. In addition, it may retrieve heuristic or historical knowledge as contexts through nearest-neighbor indexing to augment the learning process.

\noindent
\textbf{(2) Device Allocation.}
Whether DL inference computation may benefit from hardware accelerators such as GPU depends on many factors. According to our study of decision forest inference~\cite{guan2023comparison}, we found that for inference queries that involve simple models and small datasets, the overheads of moving data from host memory to GPU memory could outweigh the performance benefits brought by GPU acceleration. The resource management of the envisioned system should intelligently allocate devices like CPU and GPU to different inference queries. 

To resolve the problem, one option is to extend the physical query optimizer to model the running of each UDF that encapsulates DL operator(s) as a producer-transfer-consumer process~\cite{shi2014mrtuner} and estimate the overall latency accordingly depending on how the CPU-GPU data transfers are overlapped with CPU/GPU processing. Such an approach will be novel because existing UDF optimization works~\cite{zou2021lachesis, palkar2017weld, crotty2015tupleware} focus on logical optimization.

\vspace{5pt}
\section{Storage Co-Optimization}
\label{sec:storage}

\noindent
\textbf{Challenge 3.} How to manage the storage and caching of relational, tensor, and vector data in a unified way? 

The inference queries need to manipulate both relational data and tensor/vector data and we focus on the following opportunities.

\noindent
\textbf{(1) Accuracy-Aware Deduplication and Compression.}
Different from relational data that must be stored exactly, some errors/approximations in the tensor/vector data may not significantly affect the downstream application accuracy. In addition, similar tensor/vector data may be deduplicated approximately to reduce storage costs and memory footprint~\cite{DBLP:journals/pvldb/ZhouCDMYZZ22, schleich2023optimizing}. %

Therefore, in RDBMS, the caching of model data should facilitate error-bounded compression and deduplication. For example, the weights that are approximately shared by multiple tensors or the embedding vectors that are frequently accessed or are insignificant to the inference results should be prioritized in the caching hierarchy.

Moreover, the RDBMS should facilitate accuracy-aware query optimization.
For example, the storage optimizer may automatically employ compression, such as pruning and quantization, to create multiple versions of the same model with different size, efficiency, and accuracy trade-offs. Then, the query optimizer may dynamically select one version of the model for runtime query processing based on the latency and accuracy requirements defined in the Service Level Agreement (SLA). %
In addition, managing DL models in RDBMS will facilitate the binding of each model and its training datasets (e.g., through the catalog component), which will facilitate model selection based on data similarity~\cite{zhou2022benchmark}.

\noindent
\textbf{(2) Novel Physical Database Design.}
\textcolor{black}{
Physical database design~\cite{agrawal2000automated, rao2002automating} in RDBMS determines the partitioning of data and selection of indexing and materialized view. In our envisioned system, this component should be refactored to consolidate existing tensor partitioning, distributing, and offloading in existing DL libraries~\cite{chen2018tvm, sheng2023high, yadav2022distal} with the RDBMS storage management for tensor relations (i.e., as in the relation-centric approach). For example, the automatic tiling of the tensors in TVM~\cite{chen2018tvm} and the distribution of the tensor blocks to multiple devices in DISTAL~\cite{yadav2022distal} could be merged with the physical database design.}

In addition, it is promising to co-partition relational data and tensor data to facilitate relation-centric processing. 
For example, the first linear algebra computation in a feed-forward neural network is to multiply the feature tensor and the weight tensor, which can be converted into a join of the feature tensor and the weight tensor followed by an aggregation~\cite{DBLP:conf/cidr/Zou21, DBLP:journals/pvldb/ZhouCDMYZZ22, jankov2019declarative, jankov2020declarative}. If the feature tensor is the output of relational processing on relational data, co-partitioning the relational data and the weight tensor will avoid the shuffling process to conduct a distributed join, as we demonstrated in our prior work~\cite{zou2021lachesis}.

\eat{

}

\section{DL Serving Techniques in RDBMS}
\label{sec:serving}

\eat{Existing works in model serving systems fall into two categories: 
\noindent
\textbf{\textit{1. Black-box approaches.}}
Popular model serving systems such as TensorFlow Serving~\cite{olston2017tensorflow}, Amazon SageMaker~\cite{liberty2020elastic},  Microsoft's Azure ML~\cite{barga2015predictive}, and Rafiki~\cite{wang2018rafiki},  deploy ML models as black boxes with knobs for tuning.
For example, many of those systems~\cite{crankshaw2017clipper, wang2018rafiki, ali2022optimizing}
dynamically optimize the batch sizes to balance the accuracy and the latency.
\noindent
\textbf{\textit{2. White-box approaches.}} %
Nexus~\cite{shen2019nexus} proposes a bin-packing algorithm to place ML layers to devices and tune batch sizes for each layer. DeepSpeed~\cite{aminabadi2022deepspeed} and FlexGen~\cite{sheng2023high} offload certain weights to the CPU and even the disk. Other techniques are not limited to early prediction based on cached (intermediate) inference results~\cite{ kumar2019accelerating, balasubramanian2021accelerating, nakandala2019incremental}, sharing operators across models~\cite{lee2018pretzel}, and approximate inference ~\cite{cai2019model, kang13blazeit, kang2017noscope, kang2020jointly}.

To discuss the following challenge, we assess the RDBMS compliance with several challenging techniques.} %

\vspace{5pt}
\noindent
\textbf{Challenge 1.} Which techniques are compliant with and should be incorporated into RDBMS to accelerate the Relation-Centric and UDF-Centric approaches, and how to incorporate these techniques?

\vspace{5pt}

\noindent
\textbf{(1) Caching of Inference Results and Contexts.}
 Although there exist solutions that cache inference results~\cite{crankshaw2017clipper, kumar2019accelerating, finamore2022accelerating}, none of these solutions consider the integration with RDBMS. One option is to leverage RDBMS 
  indexing to facilitate an inference result cache. The idea is to create a table that records the features or intermediate representations (e.g., embedding vectors) of frequent inference requests and the corresponding prediction results or contexts that are auxiliary to the prediction. We then construct nearest neighbor indexing over the features so that an inference query may efficiently retrieve results or contexts from the cache.

It is promising to leverage nearest neighbor indexing widely used in vector databases~\cite{pinecone, milvus, faiss, qdrant, vespa}, such as hierarchical navigable small world (HNSW)~\cite{malkov2018efficient}, locality sensitive hashing (LSH)~\cite{zhu2016lsh}, inverted file indexing (IVF)~\cite{sivic2003video}, and product quantization~\cite{jegou2010product}, within RDBMS to facilitate the fast retrieval of inference results.

Such integration also triggers many research opportunities. 
First, the approximate caching will make a trade-off between the accuracy and latency, and may not benefit many accuracy-critical applications. Therefore, the proposed caching mechanism and the query optimizer should be application-aware, and should dynamically recommend whether to cache data for a specific application and whether to use the cache for a specific query based on the service level agreements (SLAs) and user configurations. Second, it is important to derive error bounds for the inference result caching. One option is considering a probabilistic error bound based on Monte Carlo sampling~\cite{shapiro2003monte}. Another option is to use the exact inference result caching leveraging the hashing indexing or multi-dimensional indexing. Third, the buffer pool page replacement policy also needs to be improved to coordinate the disparate access patterns of the vector data, the relational data, and various indexes.

\noindent
\textbf{(2). Pipelining of DL Computations.}
In DL serving systems, if a model exceeds available memory, it will be partitioned into multiple operators/layers, which will be dispatched to multiple devices based on the costs of the operators and the available resources of the devices~\cite{shen2019nexus, aminabadi2022deepspeed, chen2021bring}. Those devices work in parallel, composing a pipeline. A pipeline stage at each device works in a streaming style. It continuously accepts inputs from its upstreaming stage, processes the inputs, and passes the output to the downstream stage. %

However, although the "pipelining" concepts in RDBMS also exist, it mostly refers to operator fusion, and the query processing in RDBMS  mainly relies on data parallelism. %

These two approaches represent different trade-offs. The pipelining in DL frameworks is subject to the operator size limitation, where an operator or a layer must fit the resource on the corresponding device. The Relation-Centric processing in RDBMS has no such restriction because a large-scale tensor is represented as a collection of blocks that can spill to disks or be distributed to multiple devices. However, the pipelining in DL frameworks tends to be more efficient than the RDBMS parallel processing. First, there is no need to shuffle data globally as required by database join and aggregation. Second, computations can be reused across queries without re-compilation/re-scheduling overheads.

Implementing the DL pipelining mechanism in the UDF-centric approach is feasible by breaking the model UDF into multiple fine-grained operator UDFs and deploying those UDFs on different devices following the stream processing paradigm. 
However, it is challenging to accommodate both batch and stream processing in one system~\cite{das2014adaptive}. 
Another option is to offload large-scale models to DL frameworks, provided that each operator fits memory and the data transfer between the RDBMS and the DL framework is insignificant. %

\eat{
\vspace{3pt}
\noindent
\textbf{3-3. Dynamic Batching.} Batching of model inference requests could balance the latency, throughput, and resource utilization. Each deep learning operation (e.g., a kernel) may desire a different batch size~\cite{shen2019nexus, ali2020batch, ali2022optimizing}. However, in most RDBMS, all vectorized operators fused in one pipeline stage  (i.e., a job stage in SparkSQL) usually have the same batch size. 

To address the problem, on the one hand, vectorization and pipelining logic should be enhanced so that each operator can use a different batch size, and the batch size could change over time based on the variability of the available system resources. On the other hand, the RDBMS should provide a mechanism to automatically determine the batch size for each kernel/layer, which may be integrated as part of the query optimization logic or the hyper-parameter tuning component. 

\vspace{3pt}
\noindent
Storage-related techniques, such as model compression and offloading tensors to disks, are discussed in Challenge 4 (Sec.~\ref{sec:storage}).
}

 \section{Further Discussion}
 \label{sec:discussion}

\subsection{Extension to Deep Learning Training}
DL training also consists of linear algebra operations and may also benefit from the optimization techniques we proposed. While we focus on real-time deep learning inference queries, an interesting question is whether it is feasible to extend a deep learning inference system for the corresponding training job within the same infrastructure. For the DL-centric architecture, the extension is straightforward --- the underlying DL system (e.g. PyTorch) is equipped with the automatic differentiation engine that can construct the backward propagation computation graph automatically, and then the DL system can execute it for the SGD-based training computation on the hardware runtime. For the UDF-centric architecture, the extension relies on the implementation of the UDF that should be able to integrate the functionality of the corresponding backward computation and the SGD-based optimizers. For the relational-centric architecture, how the extension should be designed and implemented is still an open question --- one potential solution is to implement the corresponding backward computation as a set of separated fine-grained UDFs corresponding to each of the fine-grained UDFs representing the forward/inference computations, and then leverage a relational view to construct the query execution plan to schedule the computation over the execution environment~\cite{tang2023auto}.

\subsection{From A Monolithic System to A Loosely Coupled Ecosystem}
We argue for a system  to unify the different types of DL processing that lies in-between a monolithic system and a loosely coupled ecosystem to integrate all representations for diverse workloads. Most processing, including relational-centric, UDF-centric, and even simple DL-centric processing, could be seamlessly integrated with the RDBMS. The system will only offload complicated ML/DL processing that bottlenecks the pipeline processing with a time constraint to existing ML/DL systems. Compared to a loosely coupled ecosystem~\cite{evadb, park2022end}, where an out-of-box optimizer analyzes user code and dispatches computations to RDBMS, vector databases, and ML systems, our envisioned system can achieve better end-to-end performance for a broad class of workloads by avoiding cross-system overheads and maximizing co-optimization capabilities. Compared to a monolithic system, our envisioned system avoids reinventing the wheels for DL and tensor computation optimization, reduces the implementation cost by leveraging existing investments in DL systems, and achieves better encapsulation and transparency between RDBMS and ML systems.

\subsection{Limitations}
{Potential drawbacks of the proposed system} include the limitation of the size and complexity of the models that can be natively supported by the RDBMS with a significant performance advantage. More complicated models, such as large language models (LLMs), may achieve better training and inference latency in specialized systems. However, RDBMS can reduce memory costs and provide better security and privacy protections, regardless of the model size and complexity. Such unique features could be attractive for many applications, such as those hosted by small businesses. In addition, although it will be challenging for the envisioned system to provide native LLM support, it can serve as a high-performance retrieving engine by allowing efficient inference queries to retrieve data for augmenting LLM inferences~\cite{guu2020retrieval, zhou2022docprompting}.

\eat{ \subsection{Other Related Works}
The ecosystem extends beyond the components discussed in this vision paper. For instance, ConnectorX~\cite{wang2022connectorx} is a high-performance database connection library designed to reduce data transfer overhead for the DL-Centric architecture. Additionally, AI-Centric database~\cite{gandhi2022tensor, DBLP:journals/pvldb/KoutsoukosNKSAI21} converts relational processing to tensor processing and executes queries in deep learning frameworks, leveraging hardware acceleration and compilation/optimization support for tensor computations. TFX~\cite{baylor2017tfx} extends the ML platform with basic data processing capabilities. Although these works are relevant, it is essential to identify and consolidate redundant components to avoid duplication of functionalities, a task that remains for future work. 
}

\noindent

\section{Validation and Evidence}

A path to implement the envisioned problem should involve two steps. 
 First, we will enhance existing RDBMS to unify the three representations. Second, we will enhance the unified system to implement proposed techniques. We validate each step in the following sections.

\label{sec:evaluation}
\subsection{Unified Inference Query Processing}
To validate the potential benefits of the proposed unified query optimization strategy, we developed a naive rule-based inference query optimizer, which adaptively selects the in-database representation for each operator based on the required memory size of the operator. If the operator's memory requirement exceeds a configurable memory limit threshold, it will choose the relation-centric representation, otherwise, it will choose the UDF-centric representation. An operator's memory requirement is estimated as the sum of the operator's input size and the output size (e.g., for a matrix multiplication operator, if the input tensors have shapes $m\times k$ and $k\times n$, the memory requirement is simply estimated as $m\times k$+$k\times n$+$m\times n$). We prototyped the simple optimizer on top of netsDB~\cite{DBLP:journals/pvldb/ZhouCDMYZZ22} which is our object-oriented RDBMS
 implemented in C++~\cite{zou2021lachesis, zou2019pangea, zou2020architecture, zou2018plinycompute}, to leverage its capability in representing a model in analyzable UDFs~\cite{zou2021lachesis}. Such capability enables our optimizer to traverse through each operator and estimate its memory requirement.

We loaded multiple feed-forward neural network (FFNN) models and convolutional neural network (CNN) models into netsDB, using the optimizer to select the in-database representation. The detailed descriptions of these models are listed in Tab.~\ref{tab:ffnn-overall-1} and Tab.~\ref{tab:cnn-overall-1}.

\begin{table}[h]
    \centering
    \scriptsize
    \caption{\label{tab:ffnn-overall-1}\small Fully Connected (FC) Models (one hidden layer)}
    \begin{tabular}{cc}
        \toprule
        Model & Number of Features/Number of Neurons (hidden layer size)/Outputs \\
        \midrule
        Fraud-FC-256&$28/256/2$ \\
        Fraud-FC-512&$28/512/2$ \\
        Encoder-FC&$76/3,072/768$ \\
        Amazon-14k-FC&$597,540/1,024/14,588$ \\
        \bottomrule
    \end{tabular}
\end{table}

\begin{table}[h]
    \centering
    \scriptsize
    \vspace{-5pt}
    \caption{\label{tab:cnn-overall-1}\small Convolutional Models (Stride size = 1 and Padding size = 0)}
    \begin{tabular}{ccc}
        \toprule
        Model & Image/Input Shape & Kernel Shape \\
        \midrule
        DeepBench-CONV1&$112\times112\times64$&$64\times64\times1\times1$\\
        LandCover&$2500\times2500\times3$&$2048\times3\times1\times1$\\
        \bottomrule
    \end{tabular}
\end{table}

We compared the in-database processing of these models to popular ML systems including TensorFlow 2.5 and Pytorch 2.1.0. All of the systems run in an AWS r4.2xlarge instance, which has eight CPU cores and $61$ gigabytes memory, and $500$ GB SSD. For our prototype, the samples to be inferred are loaded to netsDB before the testing. For TensorFlow and Pytorch, the samples are loaded from PostgreSQL using a high-performance connector called ConnectorX~\cite{wang2022connectorx}. In the baselines, all hyper-parameters, such as the batch size, are fine-tuned to be optimal. In all experiments, we set the memory threshold to $2$ gigabytes, so that if the operator's memory requirements exceed this size, the relation-centric representation will be used for the operator. 

\noindent
\textbf{Latency Reduction for Small-Scale Models}
For small-scale models that fit into the memory threshold, our optimizer chose the UDF-centric representation that encapsulates all model inference operations in a single UDF.
As illustrated in Fig.~\ref{fig:small_ffnn} and Fig.~\ref{fig:small_cnn}, the proposed architecture is able to reduce the latency for inference queries that involve small-scale models. 

Because the model inference complexity is low in these workloads, the cross-system data transfer becomes the bottleneck of the DL-centric architecture. However, the proposed in-database model serving architecture can effectively alleviate such bottlenecks by running the inference computations directly on top of the data.

\begin{figure}[h]
\vspace{-10pt}
\centering{%
   \includegraphics[width=3.4in]{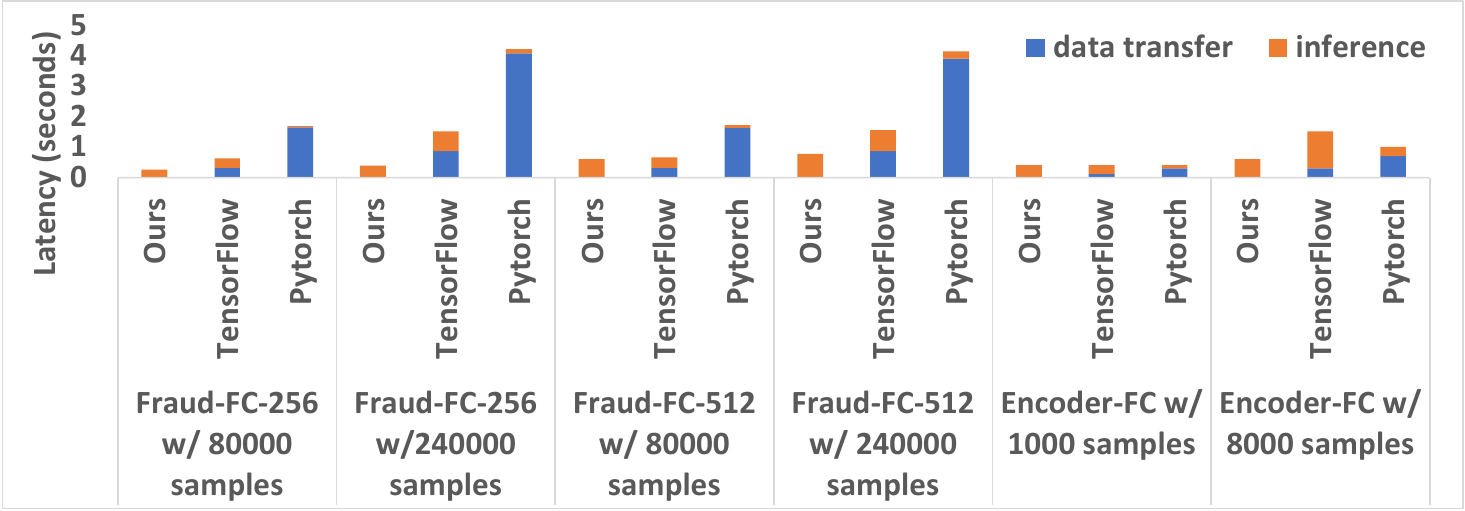}  
}
\caption{\small\label{fig:small_ffnn} Latency reduction using our rule-based optimizer for FFNN model inference over data managed by RDBMS.
}
\end{figure}

\begin{figure}[h]
\vspace{-10pt}
\centering{%
   \includegraphics[width=3.4in]{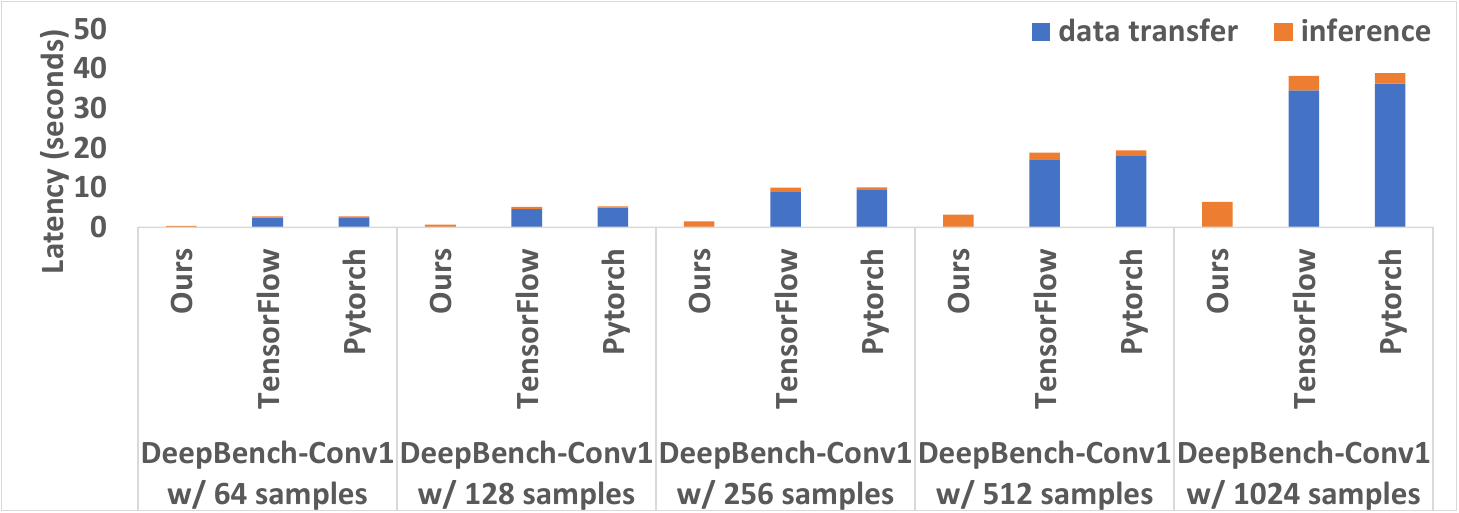}  
}
\caption{\small\label{fig:small_cnn} Latency reduction using our rule-based optimizer for CNN model inference over data managed by RDBMS.
}
\end{figure}

\vspace{5pt}
\noindent
\textbf{ OOM Error Avoidance for Large-Scale Models} 
For large-scale ML operators in which the memory requirements exceed the memory limit threshold, our optimizer chose the relation-centric representation. For example, in the AWS-14K-FC model, the tensor of the input features $X$ has a shape of $batch\_size \times 597,540$, and the weight matrix that connects the input features and the hidden layer $W$ has a shape of $1,024 \times 597,540$. Simply the weight matrix exceeds the $2$ gigabytes threshold, therefore, the operator of $X \times W^T$ would be represented in relation-centric. First, the weight matrix will be chunked into matrix blocks, and then the matrix multiplication will be converted into a join followed by an aggregation. Similarly, for the convolutional operation in LandCover, the input tensor has a shape of $batch\_size \times 2500 \times 2500 \times 3$, and the output feature map has a shape of $batch\_size \times 2500\times 2500 \times 2048$. The convolution operation will also be converted into relational operators. First, following the spatial rewriting algorithm for convolutional computation~\cite{spatial-rewrite}, each image will be flattened into a matrix $F$ of the shape $6,250,000 \times (3+1)$, and the kernel will be flattened into a matrix $K$ of the shape $2048 \times (3+1)$. Second, the convolution computation will be converted into $F \times K^T$, which will also be represented as a join followed by an aggregation. 

As illustrated in Tab.~\ref{tab:large-latency}, while the large-scale ML operator does not fit into the available memory, DL-centric architectures will throw Out-of-Memory (OOM) errors. However, the relation-centric inference processing running in the RDBMS will run at the block level, which effectively reduces the memory footprint from the scale of the tensors to the scale of the blocks. At the same time, although the entire collection of blocks cannot fit into the memory, a portion of the blocks are spilled to disk and loaded into the buffer pool (set to $20$ gigabytes) when needed. Leveraging the RDBMS buffer pool management capability, the relation-centric representation can work with resource-constrained devices. In Tab.~\ref{tab:large-latency}, we also noticed that when the ML operators of the model fit into the available memory, the DL-centric implementation relying on TensorFlow and Pytorch has a performance advantage. That's because the inference computation rather than the cross-system data transfer becomes the bottleneck in such DL-centric architecture, while our relation-centric implementation involves the additional overheads of chunking model parameter tensors into tensor blocks. %

\begin{table}[h]
    \centering
    \scriptsize
    \vspace{-5pt}
    \caption{\label{tab:large-latency} \small Latency comparison for large-scale model inferences over data managed by RDBMS}
    \begin{tabularx}{0.48\textwidth}{@{} c X X X X X @{}}
        \toprule
        Model & Batch Size & Ours & UDF centric & TensorFlow & Pytorch \\
        \midrule
        \multirow{2}*{Amazon-14k-FC} &$1000$&$58.6$&60.4&$34.6$&$\textbf{22.6}$\\
        &$8000$&$\textbf{407.2}$&OOM&OOM&OOM\\
        \midrule
        \multirow{2}*{LandCover} 
        &$1$&$36.8$& OOM&$\textbf{9.9}$ & OOM\\ 
        &$2$&$\textbf{45.2}$& OOM&OOM & OOM\\ 
        \bottomrule
    \end{tabularx}
     \vspace{-5pt}
\end{table}

\subsection{Other Representative Techniques}

\subsubsection{Model Decomposition and Push-Down} We vertically partition the Bosch dataset~\cite{guan2023comparison} that has $1.18$ millions of tuples with $968$ features into two datasets $D_1$ and $D_2$, each having $484$ features. We consider an inference pipeline that first joins $D_1$ and $D_2$ into an augmented feature dataset, denoted as $D$, through a similarity join based on the similarity of values in two columns from $D_1$ and $D_2$ respectively. These two columns are selected as having the highest correlation among pairs of columns from $D1$ and $D2$. %
We then run FFNN model over the augmented features. The model has a hidden layer of $256$ neurons and an output layer of $2$ neurons. (Therefore, the weight matrix at the first layer $W$ has a shape of $256\times 968$.) The results showed that if we decompose the matrix multiplication of $D \times W^T$ into $D_1\times W_1^T \bowtie D_2\times W_2^T$ as described in Sec.~\ref{sec:cross-optimization}, it will achieve $5.7\times$ speedup.

\eat{\subsubsection{Unified Threading}
In all results reported in Fig.~\ref{fig:small_ffnn}, Fig.~\ref{fig:small_cnn}, and Tab.~\ref{tab:large-latency}, the UDFs implemented in netsDB invoke Torch library for UDF-centric convolutional operations and Eigen library for other linear algebra operations. We found that we have to carefully tune the library threads and the netsDB threads to achieve performance gain. For example, for the UDF-centric Amazon-14K-FC implementation, by tuning the number of threads, relation-centric}

\eat{\subsubsection{Model Deduplication}
To validate that RDBMS facilitates model deduplication, we consider three models with the same architecture as Amazon-14K-FC (Tab.~\ref{tab:ffnn-overall-1}), using double precision weights. They share the weight matrix at the first fully connected layer, which has the shape of $1,024 \times 597,540$. This layer is stored as a shared set in netsDB, and it accounts for $4.8$ gigabytes of storage space. Each model also specializes the weight matrix at the second fully connected layer, with a shape of $14,588\times 1,024$ through transfer learning. Each specialized weight matrix only accounts for $0.2$ gigabytes of storage space. Therefore, with deduplication of the shared layer, the overall required storage space is reduced from $15$ gigabytes to $5.4$ gigabytes. 
In addition, we also observed up to \textbf{$1.2\times$}  speedup, because of the $40\%$ improvement in cache hit ratio due to the reduced memory footprint. 
}

\subsubsection{Caching of Inference Results}
In our preliminary experiments, we've found using Faiss' HNSW as indexing to cache the inference result, we can accelerate the inference latency for a simple CNN model with two convolutional layers (the first layer has $32$ $3\times3$ kernels, and the second layer has $16$ $3\times3$ kernels) and two fully connected layers with $64$ and $10$ neurons respectively, by $10.3\times$ speedup, with a significant accuracy reduction from $98.75\%$ to $93.65\%$. We also evaluated this method using a simple FFNN model that has four fully connected layers with $128$, $1024$, $2048$, and $64$ neurons respectively, on the MNIST dataset, and observed a speedup by $7.3\times$  with accuracy dropped from $97.74\%$ to $95.26\%$. These results motivate an adaptive inference result caching strategy which decides whether to cache the inference results by estimating a probabilistic error bound using techniques such as Monte Carlo sampling and checking whether the error bound is acceptable to the application's SLA.

\section{Conclusions}

In this work, we discuss how the RDBMS should be upgraded to support various model-serving applications that rely on RDBMS for data management. First, we identify three pivotal architectures: DL-centric, UDF-centric, and relation-centric, and we argue that these approaches as well as the middle ground among them should be unified to facilitate the serving of deep learning models on relational data at different scales. We further identify challenges and opportunities in existing RDBMS systems for achieving unified optimization. In summary, we argue for:

\noindent
$\bullet$ A novel unified IR and novel adaptive optimizer for co-optimizing inference computations and relational computations at a fine-grained style so that any subgraph in the inference query IR has the flexibility to be scheduled as DL-centric, UDF-centric, or relation-centric. Such an optimizer will open opportunities for identifying new query graph transformation rules, such as model decomposition and push-down.

\noindent
$\bullet$ Coordination of computational and memory resources for DB, DL runtimes, and lower-level runtimes that support DL libraries to unify the resource management for the future RDBMS ecosystem.

\noindent
$\bullet$ Renovating the storage, caching, and indexing of tensors/vectors to reduce storage costs, considering data placement in GPU, and facilitating accuracy-aware query optimization.

\noindent
$\bullet$ Integrating compliant model serving techniques such as caching of inference results into RDBMS and offloading non-compliant techniques to DL runtimes.

\begin{acks}
The work is sponsored by the National Science Foundation (NSF) CAREER Award (Number 2144923), the IBM Academic Research Award, the Amazon Research Award, and a U.S. Department of Homeland Security (DHS) Award (Number 17STQAC00001-07-00). Kanchan Chowdhury's work is supported by Prof. Mohamed Sarwat. 
\end{acks}

\section{Disclaimer}

The views and conclusions contained in this document are those of the authors and should not be interpreted as necessarily representing the official policies, either expressed or implied, of the U.S. Department of Homeland Security.

\bibliographystyle{ACM-Reference-Format}
\bibliography{refs}

\end{document}